\documentclass[prb,twocolumn,aps,showpacs]{revtex4}
\usepackage{graphicx}%
\usepackage{dcolumn}
\usepackage{amsmath}
\usepackage{amssymb}
\usepackage{epsfig}
\usepackage[latin1]{inputenc}
\usepackage{amsfonts}
\include{biblio}
\usepackage{color}
\usepackage{makeidx}
\begin{document}

\title{Attractive Kane-Mele-Hubbard model at half filling: phase diagram and Cooperon condensation}
\author{Zlatko Koinov  }\affiliation{Department of Physics and Astronomy,
University of Texas at San Antonio, San Antonio, TX 78249, USA}
\email{Zlatko.Koinov@utsa.edu}
 \begin{abstract}
   Recently, the attractive Kane-Mele-Habbard (KMH) model on a honeycomb lattice  at half filling has been studied in two papers: PRB 99, 184514 (2019) and  PRB 94, 104508 (2016). The authors of the first one presented the phase diagram which interpolates the trivial and non-trivial topological states. However, the next-nearest-neighbor (NNN) hopping term has been neglected, although it is several orders of magnitude stronger than the internal spin-orbit coupling. We use the mean-field approximation to derive the phase diagram of the attractive KMH model with NNN hoping at half filling. The phase diagram without and the phase diagram with NNN hopping are significantly different in the non-trivial topological region.

  The possibility to have superconducting instability in the attractive KMH model has been analyzed in the second paper within the T-matrix approximation. The question that naturally arises here is about the contributions due to the bubble diagrams, which are included in the Bethe-Salpeter (BS) equation, but neglected by the T-matrix approximation. To answer this question, we apply the BS formalism to calculate the slope of the Goldstone mode and the corresponding sound velocity. We found $4\%$ difference between the values of the sound velocity provided by the T-matrix approximation and the BS equation. This small difference confirm previously reported result that close to the phase transition boundary the bubble-diagram contributions are not important.
     \end{abstract}\pacs{71.10.Fd, 05.30.Rt, 37.10.Jk, 73.43.-f }
\maketitle
\section{Introduction}

  The present-day experiments with ultracold atoms in optical lattices allow us to simulate both the Haldane's model [\onlinecite{H2}] and the situation [\onlinecite{Exp1,Exp2}] consider by Kane and Mele (KM).  The Haldane's model [\onlinecite{Hal,H1}] is a tight-binding representation of electron motion on a honeycomb lattice in the presence of a magnetic field, which vector potential has the full symmetry of the lattice and generates a magnetic field with zero total flux through the unit cell. It was pointed out by Haldane  that in 2D honeycomb lattice the topological ordering requires time reversal symmetry breaking. Because of the zero magnetic flux through each unit cell, the phase accumulated through a nearest neighbor hopping vanishes, whereas the phase accumulated through next-nearest-neighbor (NNN) hopping is nonzero. This extra phase breaks the time-reversal symmetry. However, the electron spin is not included in the Haldane's model.

  It is known that the spin-orbit coupling preserves the time-reversal symmetry, but the spin-orbit effects can be used to get topological insulators. The first example was the  KM Hamiltonian for the electrons in a graphene [\onlinecite{KM1,KM2,WHZ}], which consists of two copies of the Haldane's model, one for spin-up electrons and one for spin-down electrons. In the KM model each spin component breaks time-reversal symmetry, but the time reversal symmetry is restored when taking two copies with different signs for the spin together.

   In a recent paper [\onlinecite{Lambda}], the phase diagram of the attractive Kane-Mele-Hubbard (KMH) model at half filling has been obtained as a function of a tuning parameter $x=3\sqrt{3}\lambda/(m_{AB}+3\sqrt{3}\lambda)$. Here $\lambda$ is the strength of intrinsic spin-orbit (ISO) coupling,  and $m_{AB}$ is sublattice potential. However, the NNN hopping term has been neglected, although it is several orders of magnitude stronger than the ISO coupling.

\begin{figure}
	\centering\includegraphics[scale=0.35]{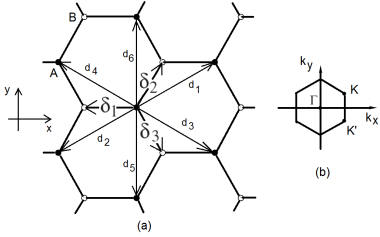}
	\caption{
Honeycomb lattice (a) and its Brillouin zone (b).}\label{F1}
\end{figure}

   In this paper, we have presented the phase diagram of the attractive KMH model with NNN hoping at half filling as a function of a tuning parameter  $x=3\sqrt{3}\lambda/(3\sqrt{3}\lambda+3t'+m_{AB})$, where $t'$ is the NNN hoping amplitude. It is shown that in the mean-field approximation we have two gap equations for $\Delta_A$ and $\Delta_B$, instead of a single gap $\Delta$ when NNN hopping is neglected. We shall discuss the case of Fermi (spin-$1/2$) atoms loaded into honeycomb optical lattice, but the results are also valid for the tight-binding description of electrons in a graphene as well. Our  tight-binding Hamiltonian  $H=H_{KM}+H_{NNN}+H_U$ includes the KM terms, as well as the NNN hopping and the onsite attractive Hubbard interaction, where

 \begin{equation}\begin{split}&H_{KM}=-t\sum_{<\textbf{i,j}>}\left[\Psi^\dag_{\textbf{i}+\textbf{j}}\Psi_{\textbf{i}}+H.c.\right]+\\
&H_{ISO} -m_{AB}\sum_\textbf{i}\left(\Psi^\dag_{\textbf{i}_A}\Psi_{\textbf{i}_A}-\Psi^\dag_{\textbf{i}_B}\Psi_{\textbf{i}_B}\right)
,\label{HKM}\end{split}\end{equation}
  The first term in (\ref{HKM}) takes into account the possibility for nearest-neighbor hopping. $H_{ISO}$ represents the ISO interaction, which  originates from the hybridization of the higher angular
momentum orbit and it exerts opposite magnetic fields upon electrons with opposite spin polarizations. The third term in (\ref{HKM})  describes the possible energy offset between sites of A and B sublattices. $t$ is the nearest-neighbor  hopping amplitudes,  $m_{AB}$ is the energy offset parameter, and $\Psi^\dag_{\textbf{i}}=(\psi^\dag_{\textbf{i},\uparrow},\psi^\dag_{\textbf{i},\downarrow})$ is the creation operator for spin-up and spin-down fermions at
site $\textbf{i}$. For hopping to the nearest neighbor sites the vectors $\textbf{j}$ in terms of lattice constant $a=1$ are: $\delta_1=(1/2,\sqrt{3}/2)$, $\delta_2=(1/2,-\sqrt{3}/2)$, $\delta_3=(-1,0)$ are shown in Fig. \ref{F1}a. The NNN hopping term is $H_{NNN}=-t'\sum_{<<\textbf{i,j}'>>}\left[\Psi^\dag_{\textbf{i}+\textbf{j}'}\Psi_{\textbf{i}}+H.c.\right]$.  For hopping to the next nearest neighbor sites the vectors $\textbf{j}'$ are $d_{1,2}=\pm(3/2,\sqrt{3}/2)$,  $d_{3,4}=\pm(3/2,-\sqrt{3}/2)$, and  $d_{5,6}=\pm(0,\sqrt{3})$. The attractive Hubbard interaction is described by $H_U=-U\sum_{\textbf{i}}\widehat{n}_{\textbf{i},\uparrow}\widehat{n}_{\textbf{i},\downarrow}$, where $U>0$.

The KM Hamiltonian along with the NNN hopping term are represented by the following $4\times 4$ matrix in the momentum space on the basis of the four-component wave function $\Psi^\dag=(\psi^\dag_{A,\textbf{k},\uparrow},\psi^\dag_{A,\textbf{k},\downarrow},\psi^\dag_{B,\textbf{k},\uparrow},\psi^\dag_{B,\textbf{k},\downarrow})$:
\begin{widetext}
\begin{equation}\widehat{H}(\textbf{k})=
\left(%
\begin{array}{cccc}
 Z(\textbf{k})+ h_{1}(\textbf{k})-m_{AB}&0&h(\textbf{k})&0\\
   0&-Z(\textbf{k})+h_{1}(\textbf{k})-m_{AB}&0&h(\textbf{k})\\
    h^*(\textbf{k})&0&-Z(\textbf{k})+h_{1}(\textbf{k})+m_{AB}&0\\
     0&h^*(\textbf{k})&0&Z(\textbf{k})+h_{1}(\textbf{k})+m_{AB}\\
\end{array}%
\right),\label{KM}\end{equation}
\end{widetext}
where  $h(\textbf{k})=-t\left[\cos(k_x)+2\cos\left(\frac{k_x}{2}\right)\cos\left(\frac{\sqrt{3}k_y}{2}\right)\right]+\imath t\left[\sin(k_x)-2\sin\left(\frac{k_x}{2}\right)\cos\left(\frac{\sqrt{3}k_y}{2}\right)\right]$,
$h_1(\textbf{k})=-2t'\left[\cos(\sqrt{3} k_y)+2\cos(\frac{3}{2}k_x)\cos(\frac{\sqrt{3}}{2}k_y)\right]$,
 $Z(\textbf{k})=-2\lambda \left[\sin(\sqrt{3}k_y)-2\cos\left(\frac{3k_x}{2}\right)\sin\left(\frac{\sqrt{3}k_y}{2}\right)\right]$, and $\lambda$ is the strength of the ISO interaction.
The eigenvalues of (\ref{KM}) are $\Omega_{1,2}(\textbf{k})=h_1(\textbf{k})\pm\sqrt{\left[m_{AB}- Z(\textbf{k})\right]^2+|h(\textbf{k})|^2}$, and $\Omega_{3,4}(\textbf{k})=h_1(\textbf{k})\pm\sqrt{\left[m_{AB}+ Z(\textbf{k})\right]^2+|h(\textbf{k})|^2}$.

\begin{figure}
	\centering\includegraphics[scale=0.35]{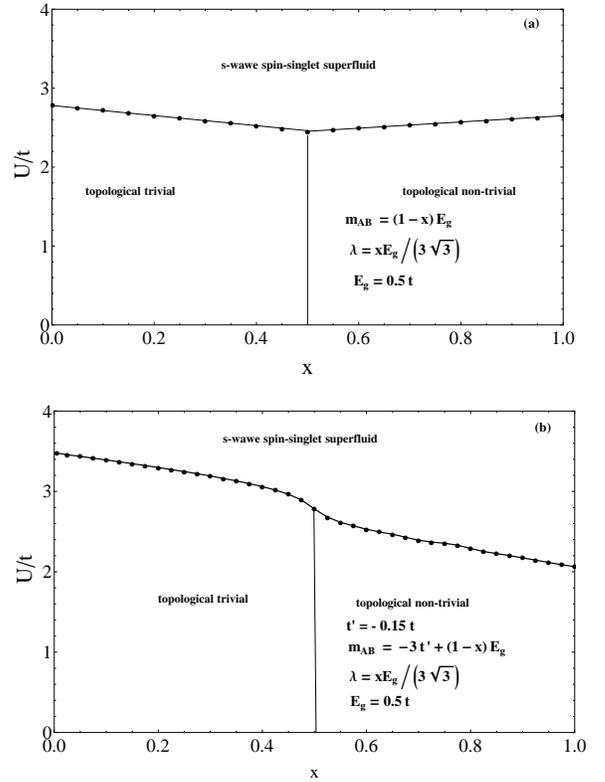}
	\caption{
Phase diagram of Kane-Mele-Hubbard model without (a), and (b) with a NNN hoping  term as functions of the tuning parameter $x$. }\label{F2}
\end{figure}
 The authors of Ref. [\onlinecite{Lambda}] pointed out that depending on the value of the  parameter $0\leq x \leq1$, the system can be taken across
the topological phase transition. For $x<1/2$ we have a topological trivial insulator, while for $x>1/2$ the system is in topological non-trivial insulator state. The  phase diagram of the KMH model, reported in Ref. [\onlinecite{Lambda}], was calculated within the self-consistent Bogoliubov- de Gennes theory using a supercell with six sites. The corresponding system parameters are $\Lambda=(1-x)E_g$, $3\sqrt{3}\lambda=xE_g$, and $E_g=t/2$.

 In Sec. II, we apply the mean-field approximation to obtain the matrix elements of the single-particle Green's functions of the KMH model with NNN hopping term  at half filling. The above-mentioned supercell approach is more complicated than the  mean-field approximation, but as can be seen from Fig. \ref{F2}a in this paper, our approach reproduces the corresponding phase diagram (Fig. 2a in Ref. [\onlinecite{Lambda}]). It turns out that the  matrix elements of the mean-field  single-particle Green's function  in the momentum space, proportional to $<\Psi_{A\textbf{k}\downarrow} \Psi_{A-\textbf{k}\uparrow}>$ and  $<\Psi_{B\textbf{k}\downarrow} \Psi_{B-\textbf{k}\uparrow}>$, are different, and therefore, we have to introduce two gaps, $\Delta_A$, and $\Delta_B$. If the NNN hopping is neglected, we have $\Delta_A=\Delta_B=\Delta$.

When the NNN hopping is included, we have $m_{AB}=-3t'+(1-x)E_g$, and $3\sqrt{3}\lambda=xE_g$. At $3t'+m_{AB}=3\sqrt{3}\lambda$ the the single-particle gap does close at $K=\left(\frac{2\pi}{3a},\frac{2\pi}{3\sqrt{3}a}\right)$ while the mass of the other bands at $K'=\left(\frac{2\pi}{3a},-\frac{2\pi}{3\sqrt{3}a}\right)$ remains constant throughout the transition for all
values of $x$, and viceversa. As in Ref. [\onlinecite{Lambda}],  the ground state of the KM Hamiltonian with NNN hopping is topological nontrivial (or topological  trivial), when the parameter $x=3\sqrt{3}\lambda/(3\sqrt{3}\lambda+3t'+m_{AB})$ is $x>1/2$ (or $x<1/2$).

Our next goal is to examine the results about possible Cooperon condensation, discussed in Ref. [\onlinecite{T}] within the T-matrix approximation. The model Hamiltonian, used in Ref. [\onlinecite{T}], corresponds to $x=1$ and $t'=0$.  Due to the attractive onsite interaction the system becomes unstable against the formation of a s-wave spin-singlet superfluid ground state. To the best of our knowledge, the superfluidity of fermion atoms in honeycomb optical lattice has been examined only in  the above mentioned paper.  According to the T-matrix  approximation,  the excitation spectrum of
collective modes  $\omega(\textbf{Q})$ was derived by calculating the roots of the following secular $2\times 2$  determinant:
\begin{equation}det\left|%
\begin{array}{cc}
U^{-1}-\Pi_{AA}(\omega,\textbf{Q})&-\Pi^*_{AB}(\omega,\textbf{Q})\\
  -\Pi_{AB}(\omega,\textbf{Q})&U^{-1}-\Pi_{BB}(\omega,\textbf{Q})
\end{array}%
\right|=0,\label{2x2}\end{equation}
where
\begin{equation}\Pi_{\xi\zeta}
 =\int
\frac{d\Omega}{2\pi}
 \frac{d^2\textbf{k}}{(2\pi)^2}G^{KM}_{\xi\downarrow,\zeta\downarrow}
 \left(\textbf{k}+\textbf{Q},\Omega+\omega\right)G^{KM}_{\xi\uparrow,\zeta\uparrow}(\textbf{k},
 \Omega).\nonumber\end{equation}
 Here $\xi,\zeta$ are the sublattice indices, and $G^{KM}_{\xi\sigma,\zeta\sigma}(\textbf{k},
 \omega)$ is the Fourier transforms of the KM single-particle Green's function $G^{KM}_{\xi \textbf{i}\sigma,\zeta\textbf{j}\sigma}(t-t')=-<\widehat{T}_t\{\psi_{\xi \textbf{i}\sigma}(t)\psi^\dag_{\zeta \textbf{j}\sigma'}(t')\}>$. It is worth mentioning that the T-matrix approximation, also known as the ladder approximation to the Bethe-Salpeter (BS) equation,  consists of the sum of ladder diagrams in the perturbation expansion in terms of $U$ where the corresponding single-particle KM Green's functions  are independent on the interaction. The question that naturally arises here is about the contributions due to the bubble diagrams, neglected by the T-matrix approximation.

 To answer the above question, in Sec. III, we derive the BS equation by employing the Hubbard-Stratonovich transformation (HST). If no approximations were made in evaluating the corresponding functional integrals, it would not matter which of the possible HST is chosen. When approximations are taken, the final result depends on a particular form chosen. A possible approximation is to introduce
the energy gap as an order parameter field, which
allows us to integrate out the fermion fields and to arrive
at an effective action. Next steps are to consider the
state, which corresponds to the saddle point of the effective
action, and to write the effective action as a series
in powers of the fluctuations and their derivatives. The
exact result can be obtained by explicitly calculating the
terms up to second order in the fluctuations and their
derivatives. This approximation, known as the Gaussian
approximation, has been employed in the case of square geometry [\onlinecite{GA}],  but to the best of our knowledge, it has never been used in the case of honeycomb lattice. In our approach, the quartic terms are transformed to quadratic forms by introducing a  boson
field which mediates the interaction of fermions. This assumption is similar to the situation in quantum electrodynamics, where the photons mediate the interaction of electric charges, and it allows us to derive the Schwinger-Dyson (SD) equation for the poles of the single-particle Green's function, as well as the BS  equation in the generalized random phase approximation (GRPA) for the poles of the two-particle Green's function.  In the GRPA, the single particle excitations are replaced with those obtained by diagonalizing the Hartree-Fock (HF) mean-field Hamiltonian, while the collective modes are obtained by solving the BS equation in
which the single-particle Green's functions are calculated
in HF mean-field approximation, and the BS kernel is obtained by
summing ladder and bubble diagrams.

We have calculated the slope of the low-energy (Goldstone) mode and the corresponding sound velocity at half filling, using the same system parameters as in Ref. [\onlinecite{T}]. We found that the T-matrix approximation is a good approximation because the sound velocity in the direction toward  point $M$, calculated within the T-matrix approximation, is about $4\%$ less than the result by employing the BS equation.
  \section{Single-particle dispersion in the mean-field approximation}
  In the presence of an  onsite attractive interaction between the fermions,  the fermion atoms form bound (Cooper) pairs. As a result, the system becomes unstable against the formation of a s-wave spin-singlet superfluid ground state. At low energies the system admits an effective description in terms of massless Dirac fermions, therefore, in a honeycomb optical lattice we have a possibility to observe a superfluidity of Fermi atoms with the Dirac spectrum. We restricted our calculations to half-filling ($\mu=0$), where the particle-hole symmetry takes place.
  We further assume that the BCS mean-field order parameters are real constants, i.e. $\Delta_{A(B)}=U<\psi_{A(B),-\textbf{k},\downarrow}\psi_{A(B),\textbf{k},\uparrow}>$.  When the attractive Hubbard interaction is taken into account,
 the KM basis of the four-component wave function $\Psi^\dag=(\psi^\dag_{A,\textbf{k},\uparrow},\psi^\dag_{A,\textbf{k},\downarrow},\psi^\dag_{B,\textbf{k},\uparrow},\psi^\dag_{B,\textbf{k},\downarrow})$
 becomes a  basis of the eight-component wave function  $\Psi^\dag= (\psi^\dag_{A,\textbf{k},\uparrow},\psi^\dag_{B,\textbf{k},\uparrow},\psi_{B,-\textbf{k},\downarrow},\psi_{A,-\textbf{k},\downarrow},\psi^\dag_{A,\textbf{k},\downarrow},\psi^\dag_{B,\textbf{k},\downarrow},\psi_{B,-\textbf{k},\uparrow},$ $\psi_{A,-\textbf{k},\uparrow})$. Thus, the generalized Hamiltonian $H=H_{KM}+H_{NNN}+H_U$ in the mean-field approximation  in the momentum space  on the basis of the eight-component wave function is represented by the following $8\times 8$ block-diagonal matrix:
 \begin{equation}\widehat{H}_{MF}(\textbf{k})=\left(%
\begin{array}{cc}
  \widehat{H}_{1}(\textbf{k})&0\\
  0&\widehat{H}_{2}(\textbf{k})\\
\end{array}%
\right),\label{H88}\end{equation}
where the corresponding $4\times 4$  blocks are defined by the following block-matrices:
\begin{widetext}
  \begin{equation}
\widehat{H}_{1}(\textbf{k})=\left(%
\begin{array}{cccc}
h_1(\textbf{k})+Z(\textbf{k})-m_{AB}& h(\textbf{k})&0&\Delta_A\\
   h^*(\textbf{k})&h_1(\textbf{k})-Z(\textbf{k})+m_{AB}&\Delta_B&0\\
   0&\Delta_B&-h_1(\textbf{k})+Z(\textbf{k})-m_{AB}&- h^*(\textbf{k})\\
    \Delta_A&0&- h(\textbf{k})&-h_1(\textbf{k})-Z(\textbf{k})+m_{AB}\\
\end{array}%
\right).\nonumber\end{equation}
 The block $\widehat{H}_{2}(\textbf{k})$ follows from $\widehat{H}_{1}(\textbf{k})$ replacing $Z(\textbf{k})$ and $\Delta_{A,B}$ by $-Z(\textbf{k})$ and $-\Delta_{A,B}$.

The eigenvalues of the $8\times 8$ Hamiltonian are as follows:
$$\omega_1(\textbf{k})=\sqrt{\Omega^2_-(\textbf{k})+\omega^2_-(\textbf{k})},\quad \omega_2(\textbf{k})=\sqrt{\Omega^2_-(\textbf{k})-\omega^2_-(\textbf{k})},\quad \omega_5(\textbf{k})=-\omega_1(\textbf{k}),\quad \omega_6(\textbf{k})=-\omega_2(\textbf{k}),$$
$$\omega_3(\textbf{k})=\sqrt{\Omega^2_+(\textbf{k})+\omega^2_+(\textbf{k})},\quad \omega_4(\textbf{k})=\sqrt{\Omega^2_+(\textbf{k})-\omega^2_+(\textbf{k})},\quad \omega_7(\textbf{k})=-\omega_3(\textbf{k}),\quad \omega_8(\textbf{k})=-\omega_4(\textbf{k}),$$
$$\Omega_{\pm}(\textbf{k})=\sqrt{(\Delta^2+\delta^2)/4+h_1^2(\textbf{k})+|h(\textbf{k})|^2+(m_{AB}\pm Z(\textbf{k}))^2},\quad \Delta=\Delta_A+\Delta_B,\quad \delta=\Delta_A-\Delta_B,$$
$$\omega^2_{\pm}(\textbf{k})=\sqrt{4h_1^2(\textbf{k})\left[|h(\textbf{k})|^2+(m_{AB}\pm Z(\textbf{k}))^2\right]+\delta^2\left[\Delta^2/4+|h(\textbf{k})|^2\right]-2\delta \Delta h_1(\textbf{k})(m_{AB}\pm Z(\textbf{k}))}.$$

\end{widetext}
The single-particle excitations in the mean-field approximation manifest themselves as poles of the Matsubara single-particle  Green's function, defined as:
 $$\widehat{G}(\textbf{k},\imath \omega_m)=\left(\imath \omega_m \widehat{1}-\widehat{H}_{MF}(\textbf{k})\right)^{-1}$$
 $$=\left(%
\begin{array}{cc}
G^{(1)}(\textbf{k},\imath \omega_m)&0\\
  0&G^{(2)}(\textbf{k},\imath \omega_m)\\
   \end{array}%
\right).$$
Here, the Matsubara fermion energies are  $\omega_m=(2\pi/\beta)(m+1/2), m=0,1,2...$,  $\beta=1/(k_BT)$, and  $k_B$ is the Boltzmann constant (throughout
this paper we have assumed $\hbar=k_B=1$). The corresponding zero-temperature Green's function
 $\widehat{G}(\textbf{k},\omega)$  is an  $8 \times 8$  matrix with elements $G^{(1,2)}_{n_1,n_2}(\textbf{k}, \omega)\quad \{n_1,n_2\}= 1,2,...8$ written in the following form:
 \begin{widetext}
   \begin{equation}\begin{split}
&G^{(1)}_{n_1,n_2}(\textbf{k}, \omega)=\frac{A^{(1)}_{n_1,n_2}(\textbf{k})}{\omega-\omega_1(\textbf{k})+\imath 0^+}+\frac{B^{(1)}_{n_1,n_2}(\textbf{k})}{\omega+\omega_2(\textbf{k})-\imath 0^+}+\frac{C^{(1)}_{n_1,n_2}(\textbf{k})}{\omega-\omega_2(\textbf{k})+\imath 0^+}+\frac{D^{(1)}_{n_1,n_2}(\textbf{k})}{\omega+\omega_2(\textbf{k})+\imath 0^+},\\
&G^{(2)}_{n_1,n_2}(\textbf{k}, \omega)=\frac{A^{(2)}_{n_1,n_2}(\textbf{k})}{\omega-\omega_3(\textbf{k})+\imath 0^+}+\frac{B^{(2)}_{n_1,n_2}(\textbf{k})}{\omega+\omega_3(\textbf{k})-\imath 0^+}+\frac{C^{(3)}_{n_1,n_2}(\textbf{k})}{\omega-\omega_4(\textbf{k})+\imath 0^+}+\frac{D^{(2)}_{n_1,n_2}(\textbf{k})}{\omega+\omega_4(\textbf{k})+\imath 0^+}.\label{GF}
\end{split}\end{equation}
 The functions $A^{(1,2)}_{n_1,n_2}(\textbf{k}), B^{(1,2)}_{n_1,n_2}(\textbf{k}), C^{(1,2)}_{n_1,n_2}(\textbf{k})$ and $D^{(1,2)}_{n_1,n_2}(\textbf{k})$ can be numerically calculated by inverting the matrix $\left(\imath \omega_m \widehat{1}-\widehat{H}_{MF}(\textbf{k})\right)$.

 The momentum distribution for the spin components $n_{\uparrow(\downarrow)}(\textbf{k})$ can be evaluated using the
 corresponding elements of the $8\times 8$ Green's function matrix:
$$n_{\uparrow}(\textbf{k})=<\psi^\dag_{A,\textbf{k},\uparrow}\psi_{A,\textbf{k},\uparrow}>+<\psi^\dag_{B,\textbf{k},\uparrow}\psi_{B,\textbf{k},\uparrow}>
     =\beta^{-1}\sum_{\imath\omega_m}\left[G_{11}(\textbf{k},\imath \omega_m)+G_{22}(\textbf{k},\imath \omega_m)\right],$$
 $$n_{\downarrow}(\textbf{k})=<\psi^\dag_{A,\textbf{k},\downarrow}\psi_{A,\textbf{k},\downarrow}>+<\psi^\dag_{B,\textbf{k},\downarrow}\psi_{B,\textbf{k},\downarrow}>
  =\beta^{-1}\sum_{\imath\omega_m}\left[G_{55}(\textbf{k},\imath \omega_m)+G_{66}(\textbf{k},\imath \omega_m)\right].$$

  Very similarly, one can derive a set of two gap equations $\Delta_A=U<\psi_{A,-\textbf{k},\downarrow}\psi_{A,\textbf{k},\uparrow}>=U\beta^{-1}\sum_{\imath\omega_m}G^{(2)}_{58}(\textbf{k},\imath \omega_m)$,   $\Delta_B=U<\psi_{B,-\textbf{k},\downarrow}\psi_{B,\textbf{k},\uparrow}>=U\beta^{-1}\sum_{\imath\omega_m}G^{(2)}_{67}(\textbf{k},\imath \omega_m)$ which at a zero temperature assume the form:
 \begin{equation}
\Delta_A=U\sum_{\textbf{k}\in BZ}\left[B^{(2)}_{58}(\textbf{k})+D^{(2)}_{58}(\textbf{k})\right],\quad
\Delta_B=U\sum_{\textbf{k}\in BZ}\left[B^{(2)}_{67}(\textbf{k})+D^{(2)}_{67}(\textbf{k})\right].\label{gap}
\end{equation}
\end{widetext}

\begin{figure}
	\centering\includegraphics[scale=0.35]{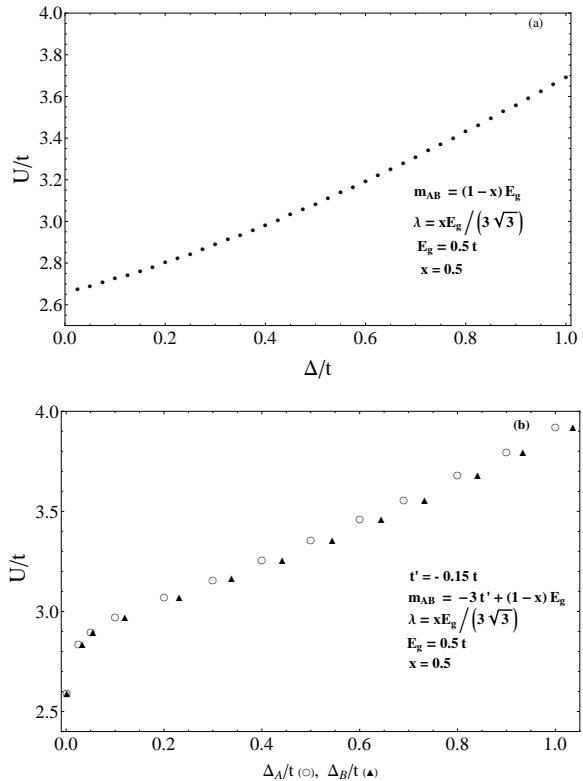}
	\caption{
  The Hubbard attractive interaction $U$ vs. the s-wave gap (gaps)  calculated for $x=1/2$: (a) $t'=0$, (b) $t'=-0.15t$. }\label{F3}
\end{figure}

 Our next step is to solve the gap equations (\ref{gap}) assuming $t'=-0.15t$ (the same value has been used previously in [\onlinecite{U1}]). At $x=1/2$, i.e. $3t'+m_{AB}=3\sqrt{3}\lambda$, the band structure is at a topological phase transition because the Dirac bands at points $K$ or $K'$ being massless (at $x=1/2$,  we have $\Omega_4(\textbf{k}_K)=0$ and $\Omega_2(\textbf{k}_{K'})=0$ at points $K=(2\pi/(3a),2\pi/(3\sqrt{3}a)$ and $K'=(2\pi/(3a),-2\pi/(3\sqrt{3}a)$). Let us assume that $\Delta_B=\alpha \Delta_A$, so we use Eqs. (\ref{gap})  to obtain an equation, $\alpha=f(\alpha,\Delta_A)$, for $\alpha$ and $\Delta_A$ with
 \begin{equation} f(\alpha,\Delta_A)=\frac{\sum_{\textbf{k}\in BZ}\left[B^{(2)}_{67}(\textbf{k})+D^{(2)}_{67}(\textbf{k})\right]}{\sum_{\textbf{k}\in BZ}\left[B^{(2)}_{58}(\textbf{k})+D^{(2)}_{58}(\textbf{k})\right]} \label{f}\end{equation}
Next, we fixed the value of $\Delta_A$, and solve iteratively our equation for $\alpha$. Having  $\alpha$ and $\Delta_A$, we used  Eqs.  (\ref{gap}) to obtain the corresponding value of $U$. The results of our numerical calculations for $x=1/2$ are presented in Fig. (\ref{F3}).  As can be seen, the minimum value of the Hubbard interaction that creates non-zero  superfluid gaps is almost the same with and without the NNN hopping term. The difference between the two gaps becomes important for higher values of $U$.
\section{ GRPA for the collective modes }
  The Green's functions in the  functional-integral approach are defined by means
  of the so-called generating functional with sources for the boson and fermion  fields.
  In our problem,  the corresponding functional integrals  cannot be evaluated
exactly because the interaction part of the Hamiltonian is
quartic in the Grassmann fermion fields. A possible way to deal with this problem is to transform the quartic terms to a
quadratic forms by introducing a  boson
field which mediates the interaction of fermions. The boson field in the honeycomb lattice
has to be an eight-component boson field $A_{\alpha}(z)$ ($\alpha=1,2,...,8$)
 interacting with  eight-component fermion spinor fields
 $ \widehat{\overline{\Psi}}
(y)=(\psi^\dag_{A,\uparrow}(y)\psi^\dag_{A,\downarrow}(y)
\psi_{A,\uparrow}(y)\psi_{A,\downarrow}(y),\psi^\dag_{B,\uparrow}(y)\psi^\dag_{B,\downarrow}(y)
\psi_{B,\uparrow}(y)\\\psi_{B,\downarrow}(y))$, and
$\widehat{\Psi}(x)=\widehat{\overline{\Psi}}^\dag (x)$.
Here, we have introduced composite variables, $z=(\textbf{r}_j,v)=(j,v)$,
$y=\{\textbf{r}_i,u\}=\{i,u\}$ and
$x=\{\textbf{r}_{i'},u'\}=\{i',u'\}$, where
$\textbf{r}_{i},\textbf{r}_{i'}$ are the lattice
site vectors, and according to imaginary-time (Matsubara) formalism
the variable $u,u'$ and $v$ range from $0$ to $\beta$.

The action of this model system is assumed to
be of the following form
 $S= S^{(F)}_0+S^{(B)}_0+S^{(F-B)}$, where:
$$S^{(F)}_0=\widehat{\overline{\psi}
}(y)\widehat{G}^{(0)-1}(y;x)\widehat{\psi} (x),$$
$$S^{(B)}_0=\frac{1}{2}A_{\alpha}(z)D^{(0)-1}_{\alpha
\beta}(z,z')A_{\beta}(z'),$$ $$ S^{(F-B)}=\widehat{\overline{\psi}}
(y)\widehat{\Gamma}^{(0)}_{\alpha}(y,x\mid z)\widehat{\psi}
(x)A_{\alpha}(z).$$

 Here we use the summation-integration
convention: that repeated variables are summed
up or integrated over. The action $S^{(F)}_0$ describes the fermion  part of the system.
The  inverse Green's function
 of free fermions
$\widehat{G}^{(0)-1}(y;x)$ is given by the following   matrix:
\begin{widetext}
\begin{equation}\widehat{G}^{(0)-1}(y;x)=\sum_{\textbf{k},\omega_m}\exp\left[\imath
\textbf{k.}(\textbf{r}_i-\textbf{r}_{i'})-\omega_m(u-u')\right]
G_{n_1n_2}^{(0)-1}(\textbf{k},\imath\omega_m) , \nonumber
\end{equation}
 where   the non-interacting Green's
   function  is defined as $\widehat{G}^{(0)-1}(\textbf{k},\imath\omega_m)=\left(\imath \omega_m \widehat{1}-\widehat{H}_{0}(\textbf{k})\right)$. The non-interacting Hamilton $\widehat{H}_{0}(\textbf{k})$ is obtained from $\widehat{H}_{MF}(\textbf{k})$ with $\Delta_{A,B}=0$.

    The action $S^{(B)}_0$ describes
the boson field which mediates
the  fermion-fermion onsite interaction in the Hubbard Hamiltonian. The Fourier transform of the bare boson propagator $\widehat{D}^{(0)}$ is an $8\times 8$ matrix:
\begin{equation}
\widehat{D}_{\alpha \beta}^{(0)} (z,z')=\frac{1}{N} \sum_\textbf{k}\sum_{\omega_p}
 e^{\left\{\imath\left[\textbf{k.}\left(\textbf{r}_j-\textbf{r}_{j'}\right)
 -\omega_p\left(v-v'\right)\right]\right\}}\widehat{D}_{\alpha \beta}^{(0)}(\textbf{k}),\quad
\widehat{D}_{\alpha \beta}^{(0)}(\textbf{k})=U\left(\delta_{1,\alpha}\delta_{2,\beta}+\delta_{2,\alpha}\delta_{1,\beta}
+\delta_{5,\alpha}\delta_{6,\beta}+\delta_{6,\alpha}\delta_{5,\beta}\right)
.\label{FTD0}\end{equation}
 Here, the Matsubara boson energies are  $\omega_p=2m\pi/\beta; p=0,1,2...$,

 The interaction between the fermion and the boson
 fields is described by the action $S^{(F-B)}$.
 The bare vertex
$\widehat{\Gamma}^{(0)}_{\alpha}(y_1;x_2\mid
z)=\widehat{\Gamma}^{(0)}_{\alpha}(i_1,u_1;i_2, u_2\mid
j,v)=\delta(u_1-u_2)\delta(u_1-v)\delta_{i_1i_2}\delta_{i_1j}\widehat{\Gamma}^{(0)}_\alpha$
is a $8\times 8$ matrix
$\widehat{\Gamma}^{(0)}_\alpha=\left(%
\begin{array}{cc}
  \widehat{\Gamma}^{(A)}_\alpha & \widehat{0}  \\
 \widehat{0}& \widehat{\Gamma}^{(B)}_\alpha \\
\end{array}%
\right)$,
where the $4\times 4$ blocks $ \widehat{\Gamma}^{(A/B)}_\alpha$ are defined in terms of the Dirac matrix $\widehat{\gamma}_0$  and the  matrices
 $\widehat{\alpha}_i$ ($\widehat{\alpha}_i$ matrices also appear in superconductivity [\onlinecite{M}]):
\begin{equation}\begin{split}
&\widehat{\Gamma}^{(A)}_\alpha=\frac{1}{2}(\widehat{\gamma}_0+\widehat{\alpha}_z)\delta_{\alpha1}
+\frac{1}{2}(\widehat{\gamma}_0-\widehat{\alpha}_z)\delta_{\alpha2}+
\frac{1}{2}(\widehat{\alpha}_x+\imath\widehat{\alpha}_y)\delta_{\alpha3}+
\frac{1}{2}(\widehat{\alpha}_x-\imath\widehat{\alpha}_y)\delta_{\alpha4},\\
&\widehat{\Gamma}^{(B)}_\alpha=\frac{1}{2}(\widehat{\gamma}_0+\widehat{\alpha}_z)\delta_{\alpha5}
+\frac{1}{2}(\widehat{\gamma}_0-\widehat{\alpha}_z)\delta_{\alpha6}+
\frac{1}{2}(\widehat{\alpha}_x+\imath\widehat{\alpha}_y)\delta_{\alpha7}+
\frac{1}{2}(\widehat{\alpha}_x-\imath\widehat{\alpha}_y)\delta_{\alpha8},\\
&\widehat{\gamma}_0=\left(%
\begin{array}{cccc}
  1&0&0&0  \\
 0&1&0&0  \\
 0& 0& -1&0  \\
 0& 0& 0&-1  \\
\end{array}%
\right),\quad  \widehat{\alpha}_i=\left(%
\begin{array}{cc}
  \sigma_i & 0  \\
 0& \sigma_y\sigma_i\sigma_y \\
\end{array}%
\right), i=x,y,z.\label{Gamma0}\end{split}\end{equation}

The basic assumption in our BS formalism is that the
bound states of two Fermi atoms in an optical lattice at zero
temperature are described by the BS wave functions (BS
amplitudes). The BS amplitude determines the probability amplitude to find the first atom at the site $\textbf{i}$ at
the moment $t_1$ and the second atom at the site $\textbf{j}$ at
the moment $t_2$. The BS amplitude depends on the relative internal time $t_1 - t_2$ and on the "center-of-mass" time
$(t_1 + t_2)/2$ [\onlinecite{IZ}]. Since the boson
propagator $\widehat{D}^{(0)}(\textbf{k})$ is frequency independent, the spectrum of the collective modes will be obtained by solving the following BS equation for the
equal-time BS amplitude $\Psi^{\textbf{Q}}_{n_2,n_1}, \{n_1,n_2\}=1,2,...,8$:
 \begin{equation}\Psi^{\textbf{Q}}_{n_2n_1}=K^{(0)}\left(%
\begin{array}{cc}
  n_1 & n_3  \\
  n_2 & n_4 \\
\end{array}%
|\omega(\textbf{Q})\right)
\left[I_d\left(%
\begin{array}{cc}
  n_3 & n_5  \\
  n_4 & n_6 \\
\end{array}%
\right)+I_{exc}\left(%
\begin{array}{cc}
  n_3 & n_5  \\
  n_4 & n_6 \\
\end{array}%
\right)\right]\Psi^{\textbf{Q}}_{n_6,n_5}.\label{BSEdZ1}
\end{equation}
In the GRPA the two-particle propagator $K^{(0)}$ is written in terms of the mean-field single-particle Green's functions:
\begin{equation}K^{(0)}\left(%
\begin{array}{cc}
  n_1 & n_3  \\
  n_2 & n_4 \\
\end{array}%
|\omega(\textbf{Q})\right)\equiv K^{(0)}_{n_1n_3n_4n_2}=\int
\frac{d\Omega}{2\pi}
 \int\frac{d^2\textbf{k}}{(2\pi)^2}G_{n_1n_3}
 \left(\textbf{k}+\textbf{Q},\Omega+\omega(\textbf{Q})\right)G_{n_4n_2}(\textbf{k},
 \Omega).\label{K0}\end{equation}
The kernel of the BS equation is a sum of
the direct $I_d=\delta \Sigma^F /\delta G$ and exchange
$I_{exc}=\delta \Sigma^H /\delta G$ interactions, written as
derivatives of the Fock  and the Hartree parts of the self-energy.
This means that the BS equation and the corresponding the SD equation for the self-energy have to be solved self-consistently. In the Appendix A, we have presented an approximation which allows us to decouple the BS and SD equations, and to obtain the following expressions for the BS kernel:
\begin{equation}
I_d\left(%
\begin{array}{cc}
  n_1 & n_3  \\
  n_2 & n_4 \\
\end{array}%
\right)=-\Gamma^{(0)}_\alpha(n_1,n_3)D^{(0)}_{\alpha\beta}
\Gamma^{(0)}_\beta(n_4,n_2),
\quad I_{exc}\left(%
\begin{array}{cc}
  n_1 & n_3  \\
  n_2 & n_4 \\
\end{array}%
\right)=
\frac{1}{2}\Gamma^{(0)}_\alpha(n_1,n_2)D^{(0)}_{\alpha\beta}
\Gamma^{(0)}_\beta(n_4,n_3).\label{Sigma}
\end{equation}
Here $\Gamma^{(0)}_\alpha(n_1,n_2)$ is the corresponding matrix element of $\widehat{\Gamma}^{(0)}_\alpha$. The BS equation, written in the matrix form, is
$\left(\widehat{I}+U\widehat{Z}\right)\widehat{\Psi}=0$, where
 $\widehat{I}$ is the unit matrix, and the condition for the existence of non-trivial solution requires the $64 \times 64$ determinant  $det\left|U^{-1}\widehat{I}+\widehat{Z}\right|=0$.  By applying simple matrix algebra, the $64\times 64$ determinant can be simplified to a $20\times 20$ one of the following form
\begin{equation}det\left|%
\begin{array}{ccc}
A_{4\times 4}(\omega,\textbf{Q})&0&0\\
0&B_{4\times 4}(\omega,\textbf{Q})&0\\
&0&C_{12\times 12}(\omega,\textbf{Q})\\
\end{array}%
\right|=0.\label{20x20}\end{equation}
\end{widetext}

 The elements of the above three blocks are given in the Appendix B. Blocks $A_{4\times 4}$ and $B_{4\times 4}$ have different elements, but $det|A_{4\times 4}(\omega,\textbf{Q})|=det|B_{4\times 4}(\omega,\textbf{Q})|$. The above  $20\times 20$ determinant vanishing if $det|A_{4\times 4}(\omega,\textbf{Q})|=0$, or $det|C_{12\times 12}(\omega,\textbf{Q})|=0$. Our numerical calculations at half filling show that $det|A_{4\times 4}(\omega=0,\textbf{Q}=0| \neq 0$. This means that  the Goldstone mode dispersion within the  BS formalism is provided by the secular determinant $Det[C_{12\times 12}(\omega,\textbf{Q})|=0$.

  To compare our numerical results with the T-matrix approximation, we assume the same  system parameters as in Ref. [\onlinecite{T}]: $m_{AB}=0$, $t'=0$, $\lambda=0.1t$, and $U=2.69t$.  The gap equation provides $\Delta=0.151t$. Having the mean-field gap, we have calculated the sound velocity $u=\alpha at/\hbar$ at half filling  in the direction of point $\Gamma$ toward point $M$. The slope $\alpha$ of the linear part of the collective-mode dispersion has been calculated numerically by using three points $\omega(Q_x,Q_y=0)$ with $Q_x=0$, $aQ_x=0, 0.0025\pi$ and $0.0050\pi$. The corresponding  slope is $\alpha=1.45$, and therefore, the sound velocity becomes $u/v_F=0.97$, where we have introduced the Fermi velocity $v_F=(3/2)ta/\hbar$ in a honeycomb lattice. For the similar system parameters, the slope, obtained from Fig. (5b) in Ref. [\onlinecite{T}], is $\alpha=1.34$, that is about $4\%$ difference.
 \section{Discussion}
To summarize,  we have numerically calculated the phase diagram of the attractive KMH model with NNN hoping at half filling within the mean-field approximation. It is shown that as soonas the NNN hoping is included, we have to solve two mean-field gap equations for $\Delta_A$ and $\Delta_B$, instead of a single gap equation for $\Delta$ in the case when NNN hopping is neglected. In the second part of this paper, we have calculated the slope of the low-energy (Goldstone) mode and the corresponding sound velocity in the direction toward  point $M$ within the BS formalism. We found that the T-matrix approximation provides the sound velocity which is about $4\%$ less than the result obtained by employing the BS equation.

It is known that the Gaussian approximation also neglects the bubble diagrams, but in a square lattice the difference between the speeds of the sound calculated in the Gaussian and in the BS approximations is about $25\%$ (see Fig 10 in Ref[\onlinecite {ZS}]). To explain the small difference of $4\%$ in our numerical calculations, we refer to the system parameters: $x=1$,  $\lambda=0.1t$,  $U=2.69t$, and $\Delta=0.151t$. From the value of $\lambda$ follows that $E_g\approx 0.52t$, and therefore, the phase diagram is very close to that presented in Fig. \ref{F2}. From another point of view, the value of  $U=2.69t$ tells us that the system is very close to the topological phase transition line at $x=1$. The fact that close to the phase transition boundary the speed of sound calculated with in the Gaussian and the BS approaches is essentially the same has been previously found in a square lattice [\onlinecite{ZR}].  Thus, it is naturally to expect that away from the phase transition boundary  contributions due to the bubble diagrams will be more important.

\begin{widetext}
  \appendix{\textbf{Appendix A}}

There is one-to-one correspondence between the KMH model and our model system, which is based on  the following  Hubbard-Stratonovich transformation for
the fermion operators:
\begin{equation}\int \mu[A]\exp\left[\widehat{\overline{\psi}}
(y)\widehat{\Gamma}^{(0)}_{\alpha}(y;x|z)\widehat{\psi}(x)A_{\alpha}(z)\right]
=\exp\left[-\frac{1}{2}\widehat{\overline{\psi}}
(y)\widehat{\Gamma}^{(0)}_{\alpha}(y;x|z)\widehat{\psi}(x)
D_{\alpha,\beta}^{(0)}(z,z') \widehat{\overline{\psi}}
(y')\widehat{\Gamma}^{(0)}_{\beta}(y';x'|z')\widehat{\psi}(x')\right].\nonumber\end{equation}
 The functional measure $D\mu[A]$ is chosen to be:
$$
\mu[A]=DAe^{-\frac{1}{2}A_{\alpha}(z)D_{\alpha,\beta}^{(0)-1}(z,z')
A_{\beta}(z')},\int \mu[A] =1.$$

 According to the field-theoretical approach, the expectation value of a general operator
$\widehat{O}(u)$ can be expressed as a functional integral over the
boson field $A$ and the Grassmann fermion fields
$\widehat{\overline{\psi}}$ and $\widehat{\psi}$:
\begin{equation}<\widehat{T}_u(\widehat{O}(u))>=\frac{1}{Z[J,M]}\int
D\mu[\widehat{\overline{\psi}},\widehat{\psi},A]\widehat{O}(u)
\exp\left[J_{\alpha}(z)A_{\alpha}(z)-\widehat{\overline{\psi}}\widehat{M}\widehat{\psi}\right]|_{J=M=0},\nonumber\end{equation}
where the symbol $<...>$ means that the thermodynamic average is
made. The
functional $Z[J,M]$ is defined by
\begin{equation}
Z[J,M]=\int
D\mu[\widehat{\overline{\psi}},\widehat{\psi},A]
\exp\left[J_{\alpha}(z)A_{\alpha}(z)-\widehat{\overline{\psi}}\widehat{M}\widehat{\psi}\right],\nonumber
\end{equation}
 where the functional measure
$D\mu[\widehat{\overline{\psi}},\widehat{\psi},A]=DAD\widehat{\overline{\psi}}D\widehat{\psi}
\exp\left(S\right)$ satisfies the condition $\int
D\mu[\widehat{\overline{\psi}},\widehat{\psi},A]=1$. The quantity
$J_\alpha(z)$ is the source of the boson field. The sources
$M_{ij}(y;x)$ of the fermion fields are included in the
$\widehat{\overline{\Psi}} (y)\widehat{M}(y,x)\widehat{\Psi}(x)$ term, where $\widehat{M}(y,x)$ is an $8\times 8$ matrix:
$$\widehat{M}(y,x)=\left(%
\begin{array}{ccc}
  M_{11}(y;x)&...&M_{18}(y;x)\\
 M_{21}(y;x)&...&M_{28}(y;x)\\
 .&...&. \\
  M_{81}(y;x)&...&M_{88}(y;x)\\
\end{array}%
\right).
$$
In what follows, we introduce complex indices  $1=\{n_1,y_1\}$, and
$2=\{n_2,x_2\}$,  so in short notations we have $M_{n_1n_2}(y_1;x_2)=M(1;2)$.

  By means of the definition of the thermodynamic average, one can express all Green's functions in terms of the functional derivatives with respect to the
  corresponding  sources
 of the
generating functional of the connected Green's functions $W[J,M]=\ln
Z[J,M]$.

  The boson Green's
function is $D_{\alpha \beta}(z,z')$ is a $8\times 8$ matrix defined
as $$D_{\alpha \beta}(z,z')=-\frac{\delta^2W}{\delta
J_{\alpha}(z)\delta J_{\beta}(z')}.$$

The single-fermion Green's function $\widehat{G}(x,y)=-<\widehat{T}_u \widehat{\Psi}(x)\otimes\widehat{\overline{\Psi}} (y)>$ includes all possible   thermodynamic averages. Its matrix elements are
$G_{n_1n_2}(x_1;y_2)=-\delta W/\delta M_{n_2n_1}(y_2;x_1)$.
The Fourier
transform of the  single-particle Green's function is given by
\begin{equation}
\widehat{G}(1;2)=
\frac{1}{N}\sum_{\textbf{k}}\sum_{\omega_{m}}\exp\{\imath\left[\textbf{k.}\left(
\textbf{r}_{i_1}-\textbf{r}_{i_2}\right)-\omega_{m}(u_1-u_2)\right]\}
\widehat{G}(\textbf{k},\imath\omega_m).\nonumber\end{equation}

The two-particle
Green's function $K\left(%
\begin{array}{cc}
  n_1,x_1 & n_3,y_3  \\
  n_2,y_2 & n_4,x_4 \\
\end{array}%
\right)$ is defined as
\begin{equation}
K\left(%
\begin{array}{cc}
  n_1,x_1 & n_3,y_3  \\
  n_2,y_2 & n_4,x_4 \\
\end{array}%
\right)=K\left(%
\begin{array}{cc}
  1 & 3  \\
  2 & 4 \\
\end{array}%
\right)=\frac{\delta^2 W}{\delta M_{n_2n_1}(y_2;x_1)\delta
M_{n_3n_4}(y_3;x_4)}=-\frac{\delta G_{n_1n_2}(x_1;y_2)}{\delta
M_{n_3n_4}(y_3;x_4)} . \nonumber
\end{equation}

 The vertex function $\widehat{\Gamma}_{\alpha}(2;1 \mid
  z)$ for a given $\alpha$  is a $8 \times 8$ matrix whose elements are:
\begin{equation}
\widehat{\Gamma}_{\alpha}(i_2,u_2;i_1,u_1 \mid
v,j)_{n_2n_1}=-\frac{\delta G_{n_2n_1}^{-1}(i_2,u_2;i_1,u_1)}{\delta
J_{\beta}(z')} D^{-1}_{\beta \alpha}(z',z). \nonumber\end{equation}

Since the single-particle and the two-particle (collective) excitations manifest themselves as  poles of the corresponding Green's functions, our next step is to obtain equations of the boson and fermion Green's functions. First, we shall obtain the SD equations, and they will be used to define the fermion self-energy (fermion mass
operator) $\widehat{\Sigma}(1;2)$. The simplest way to derive the SD equations is to use the fact that the
measure $D\mu[\overline{\psi},\psi,A]$ is invariant under the
translations $\overline{\psi}\rightarrow
\overline{\psi}+\delta\overline{\psi}$ and  $A\rightarrow A+\delta
A$:
\begin{equation}
D^{(0)-1}_{\alpha
\beta}(z,z')R_\beta(z')+\frac{1}{2}Tr\left(\widehat{G}(1;2)\widehat{\Gamma}^{(0)}_{\alpha}(2;1\mid
z)\right)+J_\alpha(z)=0, \nonumber
\end{equation}
\begin{equation}
\widehat{G}^{-1}(1;2)-\widehat{G}^{(0)-1}(1;2)+\widehat{\Sigma}(1;2)+\widehat{M}(1;2)=0,
\nonumber
\end{equation}
where $R_\alpha(z)=\delta W/\delta J_\alpha(z)$ is the average boson
field. The fermion self-energy   $\widehat{\Sigma}$, is a $8\times
8$ matrix which can be written as a sum of Hartree
$\widehat{\Sigma}^H$ and Fock $\widehat{\Sigma}^F$ parts. The
Hartree part is a diagonal matrix whose elements are:
\begin{equation}
\Sigma^H(i_1,u_1;i_2,u_2)_{n_1n_2}=\frac{1}{2}
\widehat{\Gamma}_{\alpha}^{(0)}(i_1,u_1;i_2,u_2|j,v)_{n_1n_2}
D^{(0)}_{\alpha\beta}(j,v;j',v')\widehat{\Gamma}_{\beta}^{(0)}(i_3,u_3;i_4,u_4|j',v')_{n_3n_4}
G_{n_4n_3}(i_4,u_4;i_3,u_3) .\nonumber
\end{equation}

The Fock part of the fermion self-energy is given by:
\begin{equation}\begin{split}&
\Sigma^F(i_1,u_1;i_2,u_2)_{n_1n_2}=-
\widehat{\Gamma}_{\alpha}^{(0)}(i_1,u_1;i_6,u_6|j,v)_{n_1n_6}
D^{(0)}_{\alpha\beta}(j,v;j',v')\widehat{\Gamma}_{\beta}^{(0)}(i_4,u_4;i_5,u_5|j',v')_{n_4n_5}
\times\\&K\left(%
\begin{array}{cc}
  n_5,i_5,u_5 & n_3,i_3,u_3  \\
  n_4,i_4,u_4 & n_6,i_6,u_6 \\
\end{array}%
\right)G^{-1}_{n_3n_2}(i_3,u_3;i_2,u_2).\nonumber\end{split}
\end{equation}
The Fock part of the fermion self-energy depends on the two-particle
Green's function $K$; therefore the SD equations and the BS equation
for $K$ have to be solved self-consistently.

Our approach to the Hubbard model allows us to obtain exact
equations of the Green's functions by using the field-theoretical
technique, in particular, the Legendre transforms. We can go over from the functional $W[J,M]$
to a new functional $V[R, G]=W[J[R,G],M[R,G]]-J_\alpha[R,G] R_\alpha +Tr(M[R,G]G)$, such that the conjugate equations hold:
$$\frac{\delta V}{\delta R_\alpha}(z)=-J_\alpha(z); \quad \frac{\delta V}{\delta G_{n_1n_2}(1;2)}=M_{n_1n_2}(1;2). $$
 By means of the SD equations and the identity
  $$\delta(1-3)\delta(2-4)\delta_{n_1,n_3}\delta_{n_2,n_4}=\frac{\delta M_{n_1n_2}(1;2)}{\delta  M_{n_3n_4}(3;4)}=\frac{\delta M_{n_1n_2}(1;2)}{\delta  R_\alpha(z)}\frac{\delta R_\alpha(z)}{\delta  M_{n_3n_4}(3;4)}
+\frac{\delta M_{n_1n_2}(1;2)}{\delta G_{n_5n_6}(5;6)} \frac{\delta G_{n_5n_6}(5;6)}{\delta  M_{n_3n_4}(3;4)},$$
 one sees that two-particle Green's function  satisfies the BS equation
\begin{equation}
 K^{-1}\left(%
\begin{array}{cc}
  n_2,i_2,u_2 & n_3,i_3,u_3  \\
  n_1,i_1,u_1 & n_4,i_4,u_4 \\
\end{array}%
\right)= K^{(0)-1}\left(%
\begin{array}{cc}
  n_2,i_2,u_2 & n_3,i_3,u_3  \\
  n_1,i_1,u_1 & n_4,i_4,u_4 \\
\end{array}%
\right)-I\left(%
\begin{array}{cc}
  n_2,i_2,u_2 & n_3,i_3,u_3  \\
  n_1,i_1,u_1 & n_4,i_4,u_4 \\
\end{array}%
\right).\nonumber
\end{equation}
Here, $$K^{(0)}\left(%
\begin{array}{cc}
 n_2,i_2,u_2 & n_3,i_3,u_3  \\
  n_1,i_1,u_1 & n_4,i_4,u_4 \\
\end{array}%
\right)=G_{n_2n_3}(i_2,u_3;i_2,u_2)G_{n_4n_1}(i_4,u_4;i_1,u_1)$$  is
the two-particle free propagator constructed from a pair of fully
dressed generalized single-particle Green's functions. The kernel
$I=\delta\Sigma/\delta G$ of the BS equation can be expressed as a
functional derivative  of the fermion self-energy
$\widehat{\Sigma}$. Since
$\widehat{\Sigma}=\widehat{\Sigma}^H+\widehat{\Sigma}^F$, the BS
kernel $I=I_{exc}+I_d$ is a sum of
 functional derivatives of the Hartree $\Sigma^H$ and
Fock $\Sigma^F$ contributions to the self-energy:
\begin{equation}
I_{exc}\left(%
\begin{array}{cc}
 n_2,i_2,u_2 & n_3,i_3,u_3  \\
  n_1,i_1,u_1 & n_4,i_4,u_4 \\
\end{array}%
\right)=\frac{\delta\Sigma^H(i_2,u_2;i_1,u_1)_{n_2n_1}}{\delta
G_{n_3n_4}(i_3,u_3;i_4,u_4)},\quad
I_d\left(%
\begin{array}{cc}
 n_2,i_2,u_2 & n_3,i_3,u_3  \\
  n_1,i_1,u_1 & n_4,i_4,u_4 \\
\end{array}%
\right)=\frac{\delta\Sigma^F(i_2,u_2;i_1,u_1)_{n_2n_1}}{\delta
G_{n_3n_4}(i_3,u_3;i_4,u_4)}.\nonumber\end{equation}

The BS equation and
the SD equations  have to be solved self-consistently. In order to decouple them, we note that
the  identity
\begin{equation}\begin{split}&
K^{(0)}\left(%
\begin{array}{cc}
 n_2,i_2,u_2 & n_3,i_3,u_3  \\
  n_1,i_1,u_1 & n_4,i_4,u_4 \\
\end{array}%
\right)\widehat{\Gamma}_{\beta}(i_4,u_4;i_3,u_3\mid
z')_{n_4n_3}D_{\beta \alpha}(z',z)\\&= K\left(%
\begin{array}{cc}
  n_2,i_2,u_2 & n_3,i_3,u_3  \\
  n_1,i_1,u_1 & n_4,i_4,u_4 \\
\end{array}%
\right)\widehat{\Gamma}^{(0)}_{\beta}(i_4,u_4;i_3,u_3\mid
z')_{n_4n_3}D^{(0)}_{\beta \alpha}(z',z)\nonumber
\end{split}\end{equation}

allows us to rewrite the Fock term as
\begin{equation}\Sigma^F(i_1,u_1;i_2,u_2)_{n_1n_2}=-
\widehat{\Gamma}_{\alpha}^{(0)}(i_1,u_1;i_3,u_3|j,v)_{n_1n_3}
D_{\alpha\beta}(j,v;j',v')G_{n_3n_4}(i_3,u_3;i_4,u_4)\widehat{\Gamma}_{\beta}(i_4,u_4;i_2,u_2|j',v')_{n_4n_2}
.\nonumber\end{equation}
To decouple
SD and BS equations,  we replace $D$ and
$\widehat{\Gamma}$ by the free boson propagator
$D^{(0)}$  and by the bare vertex $\widehat{\Gamma}^{(0)}$,
respectively.  In this approximation the Fock term assumes the form:
\begin{equation}\begin{split}
&\Sigma_0^F(i_1,u_1;i_2,u_2)_{n_1n_2}=-
\widehat{\Gamma}_{\alpha}^{(0)}(i_1,u_1;i_3,u_3|j,v)_{n_1n_3}
D^{(0)}_{\alpha\beta}(j,v;j',v')\widehat{\Gamma}_{\beta}^{(0)}(i_4,u_4;i_2,u_2|j',v'))_{n_4n_2}
G_{n_3n_4}(i_3,u_3;i_4,u_4).
\nonumber\end{split}\end{equation}
The total self-energy is $\widehat{\Sigma}(i_1,u_1;i_2,u_2)=\widehat{\Sigma}^H(i_1,u_1;i_2,u_2)+\widehat{\Sigma}^F(i_1,u_1;i_2,u_2)$.
The  Hartree  part of
the fermion self-energy is a diagonal $8\times 8$ matrix, but in the mean-field approximation,
 the elements on
the major diagonal of $\widehat{\Sigma}^H(i_1,u_1;i_2,u_2)$ will be included into the
chemical potential. To obtain an analytical
expression for the  single-particle Green's function in the mean-field approximation, we
neglect the
frequency dependence of the Fourier transform of the Fock part of
the fermion self-energy. In this approximation, the Fock term is an $8\times 8$ matrix with  non-zero elements $\Sigma_0^F(i_1,u_1;i_2,u_2)_{n_1n_2}=\delta(u_1-u_2)\delta({r_{i_1}}-{r_{i_2}})\Delta_{A/B}$ for $\{n_1n_2\}=\{14\},\{23\},\{32\},\{41\}$ for $\Delta_{A}$ and $\{n_1n_2\}=\{58\},\{67\},\{76\},\{85\}$ for $\Delta_{B}$.

 \appendix{\textbf{Appendix B}}

 The blocks in  Eq. (\ref{20x20}) are given by the following matrices:

$$A_{4\times 4}(\omega,\textbf{Q})=\left(%
\begin{array}{cccc}
  U^{-1}-K^{(0)}_{2211} &-K^{(0)}_{6215}&-K^{(0)}_{3214}&-K^{(0)}_{7218}  \\
  -K^{(0)*}_{6215}&U^{-1}-K^{(0)}_{6655} &-K^{(0)}_{6345}&-K^{(0)}_{7658} \\
  -K^{(0)*}_{3214}&-K^{(0)*}_{6345}& U^{-1}-K^{(0)}_{3344} &-K^{(0)}_{7348} \\
  -K^{(0)*}_{7218}&-K^{(0)*}_{7658}&-K^{(0)*}_{7348}& U^{-1}-K^{(0)}_{7788}  \\
\end{array}%
\right),$$
$$B_{4\times 4}(\omega,\textbf{Q})=\left(%
\begin{array}{cccc}
  U^{-1}-K^{(0)}_{4433} &-K^{(0)}_{8437}&-K^{(0)*}_{5436}&-K^{(0)}_{4123}  \\
  -K^{(0)*}_{8437}&U^{-1}-K^{(0)}_{8877} &-K^{(0)}_{8567}&-K^{(0)}_{8127} \\
  -K^{(0)}_{5436}&-K^{(0)*}_{8567}& U^{-1}-K^{(0)}_{5566} &-K^{(0)}_{5126} \\
  -K^{(0)*}_{4123}&-K^{(0)*}_{8127}&-K^{(0)*}_{5126}& U^{-1}-K^{(0)}_{1122}  \\
\end{array}%
\right).$$
The elements of $C_{12\times 12}(\omega,\textbf{Q})=\left(%
\begin{array}{cc}
 C_{11} & C^\dag_{21}  \\
  C_{21} & C_{22}\\
\end{array}%
\right)$ will be given by four $6\times 6$ blocks:
$$C_{11}(\omega,\textbf{Q})=\left(%
\begin{array}{cccccc}
  U^{-1}+K^{(0)}_{1144}&K^{(0)*}_{1414}&K^{(0)*}_{8145}&K^{(0)*}_{5148}&0&0  \\
  K^{(0)}_{1414}&U^{-1}+K^{(0)}_{4411} &K^{(0)*}_{8415}&K^{(0)*}_{5418}&0&0 \\
  K^{(0)}_{8145}&K^{(0)}_{8415}& U^{-1}+K^{(0)}_{8855} &K^{(0)*}_{5858}&0&0 \\
  K^{(0)}_{5148}&K^{(0)}_{5418}&K^{(0)}_{5858}& U^{-1}+K^{(0)}_{5588}&0&0 \\
  0&0&0&0&U^{-1}+K^{(0)}_{2233}&K^{(0)*}_{2323}\\
    0&0&0&0&K^{(0)}_{2323}&U^{-1}+K^{(0)}_{3322}\\
\end{array}%
\right),$$
$$C_{21}(\omega,\textbf{Q})=\left(%
\begin{array}{cccccc}
  0&0&0&0&K^{(0)}_{7236}&K^{(0)}_{7326}  \\
0&0&0&0&K^{(0)}_{6237}&K^{(0)}_{6327}  \\
  K^{(0)}_{4144}&K^{(0)}_{4414}& K^{(0)}_{8445} &K^{(0)}_{5448}&-K^{(0)}_{2232}&-K^{(0)}_{2322} \\
  K^{(0)}_{8148}&K^{(0)}_{8418}& K^{(0)}_{8858} &K^{(0)}_{8588}&-K^{(0)}_{6236}&-K^{(0)}_{6326} \\
  -K^{(0)}_{1114}&-K^{(0)}_{1411}& -K^{(0)}_{8115} &-K^{(0)}_{5118}&K^{(0)}_{3233}&K^{(0)}_{3323} \\
    -K^{(0)}_{5145}&-K^{(0)}_{5415}&- K^{(0)}_{5855} &-K^{(0)}_{5558}&K^{(0)}_{7237}&K^{(0)}_{7327} \\
\end{array}%
\right).$$
$$C_{22}(\omega,\textbf{Q})=$$
$$\left(%
\begin{array}{cccccc}
  U^{-1}+K^{(0)}_{7766}& & & & &   \\
  K^{(0)}_{6767}&U^{-1}+K^{(0)}_{6677} & & H.c. & & \\
  -K^{(0)}_{7226}&-K^{(0)}_{6227}& K^{(0)}_{2222}+K^{(0)}_{4444} & & &  \\
  -K^{(0)}_{6766}&-K^{(0)}_{6676}&K^{(0)}_{6226}+K^{(0)}_{8448}& K^{(0)}_{6666}+K^{(0)}_{8888}&  &   \\
  K^{(0)}_{7336}&K^{(0)}_{6337}&2U^{-1}-K^{(0)}_{2332}-K^{(0)}_{4114}&0-K^{(0)}_{6336}-K^{(0)}_{8118}&K^{(0)}_{1111}+K^{(0)}_{3333}
  & \\
    K^{(0)}_{7767}&K^{(0)}_{76777}&-K^{(0)}_{5445}-K^{(0)}_{7227}&2U^{-1}-K^{(0)}_{6776}-K^{(0)}_{8558}&K^{(0)}_{5115}+K^{(0)}_{7337}&K^{(0)}_{5555}+K^{(0)}_{7777}\\
\end{array}%
\right).$$
\end{widetext}

\end{document}